\documentclass[authoryear,preprint,review,12pt]{elsarticle}
\usepackage{amssymb}
\usepackage{amsmath}
\usepackage[final]{changes}

\journal{Chemical Geology}

\begin{document}

\begin{frontmatter}

\title{Setting a Limit on Anthropogenic Sources of Atmospheric $^{81}$Kr through Atom Trap Trace Analysis}

\author[ANL,UofC]{J. C. Zappala}
\author[ANL]{K. Bailey}
\author[USTC]{W. Jiang}
\author[ANL]{B. Micklich}
\author[ANL]{P. Mueller}
\author[ANL]{T. P. O'Connor}
\author[Bern]{R. Purtschert}

\address[ANL]{Physics Division, Argonne National Laboratory, Argonne, Illinois 60439, USA}
\address[UofC]{Department of Physics and Enrico Fermi Institute, University of Chicago, Chicago, Illinois 60637, USA}
\address[USTC]{School of Nuclear Science and Engineering, University of Science and Technology of China, Hefei, China}
\address[Bern]{Climate and Environmental Physics Division, Physics Institute, University of Bern, CH-3012 Bern, Switzerland}

\begin{abstract}

We place a 2.5\% limit on the anthropogenic contribution to the modern abundance of $^{81}$Kr/Kr in the atmosphere at the 90\% confidence level. Due to its simple production and transport in the terrestrial environment, $^{81}$Kr (half-life = 230,000 yr) is an ideal tracer for old water and ice with mean residence times in the range of 10$^{5}$-10$^{6}$ years. In recent years, $^{81}$Kr-dating has been made available to the earth science community thanks to the development of Atom Trap Trace Analysis (ATTA), a laser-based atom counting technique. Further upgrades and improvements to the ATTA technique now allow us to demonstrate $^{81}$Kr/Kr measurements with relative uncertainties of 1\% and place this new limit on anthropogenic $^{81}$Kr. As a result of this limit, we have removed a potential systematic constraint for $^{81}$Kr-dating.

\textbf{keywords}: $^{81}$Kr; groundwater dating; anthropogenic effects; radioisotope tracers; noble gas tracers; atom trap trace analysis

\end{abstract}

\end{frontmatter}

\section{Introduction}

Since the discovery of $^{81}$Kr in the atmosphere \citep{L1969}, the geoscience community has sought to apply this noble gas isotope as an environmental tracer. It is predominantly produced by cosmic rays impinging on stable krypton isotopes in the atmosphere and decays by electron capture to $^{81}$Br with a half-life of (2.29 $\pm$ 0.11) $\times$ 10$^{5}$ years \citep{B1993}, resulting in a cosmogenic atmospheric $^{81}$Kr/Kr ratio of 6 $\times$ 10$^{-13}$ \citep{C2004}. The long half-life, along with its inert chemical nature, makes it an ideal tracer for water and ice in the range of 10$^{5}$-10$^{6}$ years, a range beyond the reach of $^{14}$C dating \citep{L2014}.

The Atom Trap Trace Analysis (ATTA) technique was developed to perform routine $^{81}$Kr analysis. ATTA is a laser-based atom counting method, free of interferences from other isotopes, isobars, atomic, or molecular species. Since the completion of the third-generation ATTA-3 system at Argonne National Laboratory (ANL) \citep{J2012} $^{81}$Kr/Kr ratios have been routinely measured in environmental samples with better than 10\% precision for a variety of applications using both water \citep{A2015} and ice samples \citep{B2014}, typically requiring samples sizes of 100 kg of water or 40 kg of ice. As we demonstrate in this paper, recent developments in ATTA have improved the precision of the technique to the 1\% level. These advances offer a unique opportunity to probe the effects of human activity on the $^{81}$Kr abundance in our atmosphere. Not only does such a measurement better our understanding of \added{the} anthropogenic impact on isotopes in the atmosphere, but it also addresses the potential systematic effects that such activity could have on dating old groundwater, especially as ATTA measurements are pushed to even higher precision.

Since the advent of human nuclear activity, we have potentially been injecting anthropogenic $^{81}$Kr into the atmosphere. Testing of nuclear devices, nuclear fuel reprocessing, and usage of medical isotopes are all potential anthropogenic sources of $^{81}$Kr. Any anthropogenic contribution of $^{81}$Kr from these sources above the precision level of current dating measurements would disturb the atmospheric baseline over the past 100 years thereby adding an unacknowledged systematic error on all $^{81}$Kr-dating measurements. When \added{the} viability of $^{81}$Kr as an environmental tracer was initially demonstrated in 1999 using accelerator mass spectrometry (AMS), no difference was measured between a sample of krypton from air before human nuclear activity began and another sample from after the end of atmospheric nuclear weapons testing \citep{C1999}. This measurement carried a one-sigma, relative uncertainty of $\sim$30\%. Subsequently, with the development of the second-generation ATTA-2 system, the same experiment was conducted with one-sigma, relative uncertainties around the 8\% level \citep{D2003}. Previous theoretical considerations of anthropogenic $^{81}$Kr estimated the effect to be at or below the 0.01\% level \citep{C1999}.

In this paper, we perform \added{two} 1\% one-sigma relative uncertainty measurements using ATTA to place a 2.5\% experimental limit on anthropogenic $^{81}$Kr in the atmosphere at the 90\% confidence level. Given the high precision of this limit, we first present a more detailed theoretical model for anthropogenic $^{81}$Kr production in the atmosphere, which sets a new upper limit of 0.15\%. We then briefly describe the technical developments that allow ATTA to measure $^{81}$Kr/Kr ratios at the 1\% level. Finally, we perform measurements of samples from the modern era and before the nuclear age at this uncertainty level and use them to place a limit on anthropogenic $^{81}$Kr.

\section{Anthropogenic $^{81}$Kr Sources}

Along with the first experimental limit placed on anthropogenic $^{81}$Kr production \citep{C1999}, several theoretical calculations were presented to estimate the expected amount of anthropogenic $^{81}$Kr in the atmosphere. Those calculations suggested less than 0.01\% contribution of anthropogenic $^{81}$Kr to the atmosphere. Here we provide an update to those calculations including an in-depth model that removes several assumptions made in the previous calculations. To determine a scale for how many $^{81}$Kr atoms would cause a 1\% contribution, we use the fact that krypton is 1.1 parts per million by volume \citep{A2005} in the atmosphere (using an atmospheric mass of 5.1 $\times$ 10$^{18}$ kg \citep{N2016}) and that the isotopic abundance of $^{81}$Kr/Kr is 6 $\times$ 10$^{-13}$ \citep{C2004} to determine that there are about $7\times10^{25}$ $^{81}$Kr atoms in the atmosphere. Thus we are searching for effects that cumulatively produce $7\times10^{23}$ or more $^{81}$Kr atoms, i.e. 1\% of the cosmogenic inventory.
 
There are three potential anthropogenic sources that can contribute $^{81}$Kr to the atmosphere as described by \citet{C1999}: decay of $^{81}$Rb produced for medical usage, release of $^{81}$Kr during nuclear fuel reprocessing, and direct and neutron-induced production of $^{81}$Kr during the period of above-ground nuclear weapons testing.

$^{81}$Rb, produced for medical applications, decays to the isomer $^{81}$Kr$^*$, and the gamma rays emitted during the decay of this excited state to $^{81}$Kr are used to image respiratory systems. The resulting $^{81}$Kr was determined to be the smallest of the anthropogenic contributions in \citet{C1999} extrapolating from an inventory based on estimated $^{81}$Rb production in the United Kingdom from 1975 - 1996. The expected contribution was on the order of 10$^{19}$ $^{81}$Kr atoms. Since there has been no significant expansion of this technique after this inventory was taken, we continue to consider this contribution as negligible for our purposes.

Nuclear fuel reprocessing, by which spent nuclear fuel rods are recycled, requires the chemical processing of these fuels rods, causing a release of trapped gaseous decay remnants. For instance, $^{235}$U or $^{239}$Pu fissions produce $^{85}$Kr and other short-lived isotopes which decay to $^{85}$Kr. Since $^{85}$Kr has a half-life of 10.76 years, the majority of it has not decayed by the time the rods are processed. At that point the trapped $^{85}$Kr is released into the atmosphere. Reprocessing has had an enormous effect on the isotopic abundance of $^{85}$Kr/Kr in the atmosphere, increasing the ratio over the natural equilibrium in the atmosphere by over four orders of magnitude in the last sixty years \citep{A2013}.

$^{81}$Kr production due to reprocessing is strongly suppressed \replaced{compared to}{versus} $^{85}$Kr. $^{81}$Kr has a substantially lower neutron-induced-fission yield and is shielded from the decay chain of neutron-rich mass 81 isobars by stable $^{81}$Br. Moreover, the natural equilibrium of $^{81}$Kr/Kr in the atmosphere is also greater than that of $^{85}$Kr/Kr due to the the longer half-life. It nonetheless remains a potential anthropogenic contribution. In \citet{C1999}, $^{85}$Kr activity determined in 1985 by \citet{v1985} is doubled to produce an estimate of $^{85}$Kr atoms produced by nuclear fuel reprocessing up to the present day. Then the ratio of the neutron-induced fission yields between $^{81}$Kr and $^{85}$Kr was used to determine the resulting number of $^{81}$Kr atoms. We follow the approach of \citet{C1999}, but consult a more recent and extensive inventory of $^{85}$Kr production in \citet{A2013} to determine a cumulative $^{85}$Kr emission up until 2002\footnote{As we will discuss in a later section, the krypton from the modern era used to compare with pre-nuclear krypton was sampled from the atmosphere no later than 2002, and thus this calculation represents the largest possible contribution that we can measure.} of 11,800 PBq, equivalent to $5.8 \times 10^{27}$ $^{85}$Kr atoms. Applying the ratio of neutron-induced fission yields from \citet{E1994} of $6.9 \times 10^{-3}$ for $^{85}$Kr and $2.4 \times 10^{-9}$ for $^{81}$Kr gives $2.0 \times 10^{21}$ $^{81}$Kr atoms. This is equivalent to a $\sim$0.003\% anthropogenic signal, which is much smaller than we expect to detect.

The largest and most complex source of anthropogenic $^{81}$Kr is the testing of nuclear devices in the atmosphere. When a nuclear device detonates, products of the fission or fusion processes are released along with cascades of neutrons. Some of the daughter isotopes from fissions can be $^{81}$Kr atoms, and thus contribute to the anthropogenic signal. However, much more importantly, the released neutrons can be captured (or scattered to lower energies and subsequently captured) by $^{80}$Kr and $^{82}$Kr atoms in the atmosphere resulting in reactions that also produce $^{81}$Kr.

Calculations in \citet{C1999} determined $^{81}$Kr production from the $^{80}$Kr(n,$\gamma$)$^{81}$Kr reaction by examining the nuclear testing contribution of $^{14}$C from $^{14}$N(n,p)$^{14}$C, which is caused by the same cascade of neutrons, and using the ratio of the thermal cross-sections between the two reactions. For $^{81}$Kr production from the $^{82}$Kr(n,2n)$^{81}$Kr reaction, a similar approach was applied, but by using the anthropogenic contribution of $^{39}$Ar from $^{40}$Ar(n,2n)$^{39}$Ar reactions and assuming the cross-sections for both reactions to be equal.

Our theoretical model takes a more direct approach of determining $^{80}$Kr(n,$\gamma$)$^{81}$Kr and $^{82}$Kr(n,2n)$^{81}$Kr reaction rates per neutron released from the different types of nuclear devices. This approach accounts for the varying cross-sections over the whole energy range of released neutrons and does not rely upon calculations made on other elements (i.e. $^{14}$N and $^{40}$Ar). We calculate these reaction rates by simulating neutrons from a nuclear weapon source at the center of a 1km radius sphere in air using the Monte Carlo N-Particle transport code (MCNP) \citep{G2013}.

We define the air in our simulation using the \replaced{breakdown}{break down} by volume given in \citet{H1996}. We also allow for the addition of 0-5\% water by volume, which replaces the other constituents proportionally and represents the varying moisture in the air given by humidity \citep{W2006}.

The spectrum of fission-produced neutrons can be approximated by a Watt spectrum \citep{S1996}, defined as 
\begin{equation}
f(E) = Ce^{\frac{-E}{a}}\sinh(bE)^{\frac{1}{2}}
\end{equation}
where $E$ is the outgoing neutron energy in MeV; $a$ and $b$ are parameters that depend on the incident neutron\added{'}s energy and the isotope undergoing fission, in units of MeV and MeV$^{-1}$, respectively; and $C$ is a normalization constant. For these calculations, we used the MCNP default values of $a$ and $b$ which represent an average over incident neutron energies and fissioning nuclei, since the values for specific neutron energies and target nuclei do not vary the results (nuclide production rates) significantly. MCNP uses continuous-energy neutron transport to determine the neutron fluence rate as a function of energy down to thermal energies, and computes nuclide production rates from the energy-dependent fluence rate and the energy-dependent cross-sections.

For fusion sources we also use the MCNP, but must consider a different neutron spectrum. In fusion weapons the bulk of the thermonuclear energy is produced by four reactions \citep{G1960}
\begin{eqnarray}
\text{D} + \text{D} &\rightarrow& ^{3}\text{He}\;(0.82 \text{MeV}) +  n \;(2.45 \text{MeV})\label{d1}\\
\text{D} + \text{D} &\rightarrow& \text{T}\; (1.01 \text{MeV}) +  \text{H}\; (3.02 \text{MeV})\label{d2}\\
\text{D} + \text{T} &\rightarrow& ^{4}\text{He}\; (3.5 \text{MeV}) +  n \;(14.1 \text{MeV})\label{t}\\
\text{D} + ^{3}\text{He} &\rightarrow& ^{4}\text{He}\; (3.6 \text{MeV}) +  \text{H}\; (14.7 \text{MeV})\label{he3}
\end{eqnarray}
where reactions (\ref{d1}) and (\ref{d2}) are chiefly used to breed helium-3 and tritium for reactions (\ref{t}) and (\ref{he3}).\footnote{There are obviously a number of other reactions, however, given their relatively lower cross-sections \citep{S2014} we can ignore them for our purposes.} In order to reach high enough temperatures for these reactions to begin, a smaller fission bomb boosted by the presence of deuterium and tritium gas is first detonated within the fusion device \citep{U2009}. We will assume that the energies and products of this smaller fission device are consumed by the fusion fuel and ignore them for the purposes of our calculation. The fusion fuel is typically lithium deuteride \citep{U2009}, which provides not only deuterium, but lithium-6 and lithium-7 that react with neutrons and breed tritium as well \citep{K1995}. Since we are only interested in situations where net neutrons are produced we consider only equations (\ref{d1}) and (\ref{t}), which produce what we call DD (deuterium-deuterium) and DT (deuterium-tritium) neutrons respectively. We consider them to be mono-energetic neutrons\footnote{They have an energy width, but it is calculable and of order $\sim$ 100 keV \citep{K1995}, which is negligible for our purposes given their high mean energies.} which have the energies shown in equations  (\ref{d1}) and (\ref{t}).

\begin{table}
\caption{Fission yields used for anthropogenic $^{81}$Kr production calculations from \citet{E1994}}
\begin{tabular}{l l l l}
\hline
Fissile Isotope & Direct $^{81}$Kr Yields & & \\
& Thermal & Fast (1 MeV) & High (14 MeV) \\
\hline
$^{235}$U  & 9.4 $\times$ 10$^{-12}$ & 5.5 $\times$ 10$^{-12}$ & 3.7 $\times$ 10$^{-9}$ \\
$^{238}$U  & n/a & 1.1 $\times$ 10$^{-13}$ & 1.4 $\times$ 10$^{-10}$ \\
$^{239}$Pu & 2.1 $\times$ 10$^{-9}$ & 4.4 $\times$ 10$^{-10}$ & 1.8 $\times$ 10$^{-8}$ \\
\hline
\end{tabular}
\label{yield}
\end{table}

Using these three source spectra (Watt, DD, DT) and our dry to humid atmosphere, we calculate $^{80}$Kr(n,$\gamma$)$^{81}$Kr and $^{82}$Kr(n,2n)$^{81}$Kr reaction rates for various situations, and then simply multiply by the total number of source neutrons in each situation. From 1945 - 1993, fission tests produced the equivalent energy of 217 Megatons of TNT and fusion tests produced 328 Megatons over 520 atmospheric tests \citep{U1993}. There are additional underground tests equivalent to 90 Megatons of TNT that we neglect due to the comparatively small interaction of neutrons with the atmosphere from such tests. For fission tests, we assume 2.5 neutrons produced per fission and 200 MeV expended per fission to determine the number of neutrons. For fusion, we assume 17.6 MeV per DT neutron and 7.3 MeV per DD neutron\footnote{The 7.3 MeV energy used is the combined total energy of equations (\ref{d1}) and (\ref{d2}) since they have comparable reaction rates.} and treat DT as 100 times more likely than DD, due to the larger cross-section \citep{G1960} and the assumption of sufficient tritium availability from lithium reactions. Note that we are considering that the entire explosive yield is generated purely from these fusion reactions, which is an overestimate, but gives a higher upper limit, which we prefer. According to \citet{U2009}, 50\% (or more) of the energy released in a thermonuclear fusion weapon is generally from the high-energy-neutron-induced fission of natural uranium (that we will consider to be purely $^{238}$U) which is packed into a shell around the whole fusion weapon, so we include that in our calculations as well. Finally, for each fission (whether from fission weapons, or this uranium shell in fusion weapons), we applied direct fission yields of $^{81}$Kr from \citet{E1994} shown in Table \ref{yield}. The error on the reaction rates is $\pm 5\%$.

\begin{table}
\caption{\small{Calculated $^{81}$Kr atom production by source due to nuclear device detonation (1945-1993)}}
\begin{tabular}{l l l l l}
\hline
Source & $^{80}$Kr(n,$\gamma$)$^{81}$Kr  & $^{82}$Kr(n,2n)$^{81}$Kr & direct fission &\% of cosmogenic $^{81}$Kr\\
\hline
Fission  & 3.5 $\times$ 10$^{22}$ & 8.5 $\times$ 10$^{17}$ & 2.9 $\times$ 10$^{19}$ & 0.050\%  \\
Fusion & & & &  \\
DT  & 4.6 $\times$ 10$^{22}$ & 1.8 $\times$ 10$^{22}$ & n/a & 0.066\% \\
DD  & 2.4 $\times$ 10$^{21}$ & n/a & n/a & 0.003\% \\
$^{238}$U  & 2.6 $\times$ 10$^{22}$ & 6.4 $\times$ 10$^{17}$ & 3.2 $\times$ 10$^{18}$ & 0.037\% \\
\hline
Total & 1.1 $\times$ 10$^{23}$ & 1.8 $\times$ 10$^{22}$ & 3.2 $\times$ 10$^{19}$ & 0.15\% \\
\hline
\end{tabular}
\label{products}
\end{table}

We provide the final results of our calculations in Table \ref{products}, aiming to achieve the highest-bounding estimate. Thus the results shown are obtained using the dry air variants (since they produce the higher yields by a factor of 1.2-1.4 against the most humid cases), and applying the $^{239}$Pu yields for neutron-induced fission in Table \ref{yield} (which are two orders of magnitude higher than $^{235}$U yields). In total, nuclear weapons testing provides a $^{81}$Kr anthropogenic signal of $\leq$0.15\%, two orders of magnitude higher than the estimate in \citet{C1999} due to our accounting for the cross-sections of the reactions over all energies and applying spectra for different devices.

As a check of the validity of our model, particularly in light of our considerable increase over the previous estimate, we also calculated reaction rates for $^{14}$N(n,p)$^{14}$C and $^{40}$Ar(n,2n)$^{39}$Ar in order to compare our model's calculations for anthropogenic $^{14}$C and $^{39}$Ar production against those values reported in \citet{C1999}. We find that, even in this highest bounding limit scenarios, the model agrees with the anthropogenic $^{14}$C and $^{39}$Ar abundances cited in \citet{C1999} within a factor of 2.

\section{ATTA Technical Developments}
\label{ATTA} 

\begin{figure}[t]
\includegraphics[width=35pc]{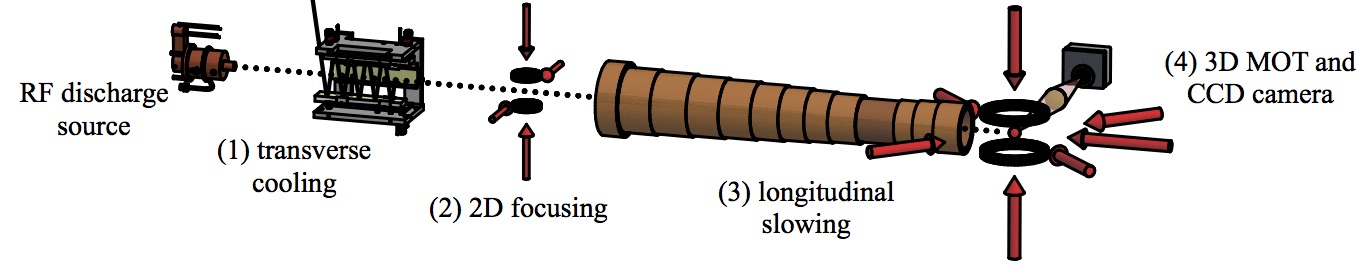}
\caption{\small{\added{Diagram of the cooling and trapping stages on the ATTA-3 beamline at ANL.}}}
\label{beamline}
\end{figure}

ATTA measures relative $^{81}$Kr/Kr ratios in a krypton gas sample by laser cooling and trapping $^{81}$Kr and $^{83}$Kr (a stable isotope of krypton) atoms from the sample and comparing the rates at which they are loaded into the atom trap. The krypton atoms are first excited by a plasma discharge to a metastable state from which a strong 811 nm cycling transition is available for cooling and trapping of the atoms. There are four stages of cooling and trapping\added{, shown in Figure \ref{beamline},} that determine the loading rate: (1) transverse cooling, (2) two-dimensional focusing using a 2D magneto-optical trap (MOT), (3) longitudinal slowing via a Zeeman slower, and (4) three-dimensional trapping using a 3D MOT. In the 3D MOT the atoms are confined at the minimum of a quadrupole magnetic field gradient and detected via laser-induced fluorescence using a sensitive CCD camera. To calibrate the efficiency we measure a standard krypton reference gas, which has a $^{81}$Kr/Kr ratio of modern air. The most recent iteration of the ATTA system at ANL, ATTA-3, is described in full detail in \citet{J2012}.

In order to reach the next level of sensitivity and place a more stringent limit on anthropogenic $^{81}$Kr, several upgrades have been implemented on the ATTA-3 system. First, we increased the excited fraction of our krypton atoms by installing an additional turbomolecular pump to compress gas into the source volume. The resulting increase in pressure within the plasma discharge produces higher excitation efficiency and increased throughput in our atom beam, without requiring an increase in sample size. Second, for the longitudinal slowing stage, we have implemented power stabilization feedback on our slowing laser after determining that drifts of $\sim$10\% in laser power can result in larger than 3\% shifts in isotope loading ratios. Third, at the detection stage, we implemented a new, ion-based detection method for the stable krypton isotope $^{83}$Kr detailed in \citet{J2014}. Finally, due to the increased speed of our measurements from the above improvements, we now measure our reference gas daily after each sample measurement, instead of only once every two weeks. This improvement removes long-term drifts in the measured isotope ratios, as large as 5\%, that can occur within the two weeks.

As a result of the source upgrade we have increased the count rate such that the typical single measurement $^{81}$Kr/Kr error, which is dominated by $^{81}$Kr statistics, has been reduced from $\sim$9\% to $\sim$3\%. All of the upgrades which have improved stabilization and precision have reduced the overall systematic errors from $\sim$5\% to $\sim$1\%.

\section{Measurement Systematics \& Procedures}
\label{procedure}

\noindent To place an improved limit on anthropogenic Kr-81 in the atmosphere, we use ATTA to measure the $^{81}$Kr/Kr isotopic ratios in two different samples: one from air before the advent of human nuclear activity (``PreAnthropogenic''), and one representing the isotopic abundances in modern air (``Modern''). Any difference in the ratio between these two samples would be interpreted as the anthropogenic contribution to our atmosphere. The PreAnthropogenic sample was prepared at the University of Bern from air in 1944 \citep{K1980}. LLC performed at the University of Bern shows the $^{85}$Kr activity of this sample to be $<$1.0 decay per minute per cubic centimeter krypton at STP (dpm/cc). The Modern sample is a commercial bottle of krypton gas purchased in June 2002 and filled at the AGA/Linde facility in Maumee, OH. Although these commercial gases are extracted from air, the exact time of sep\replaced{a}{e}ration is unknown. To ensure that this krypton gas is representative of modern air, we performed LLC at the University of Bern in March 2016 to determine the $^{85}$Kr activity. We measured an activity of 32.1 $\pm$ 1.2 dpm/cc. Using the values from \citet{A2013} and extrapolating the activity of our krypton gas backward in time, we find it to be consistent with krypton taken from the air in 2002.

As noted in Section \ref{ATTA}, the relative efficiency of our system can slowly drift over timescales on the order of one week. Thus, the $^{81}$Kr/Kr isotopic ratios of PreAnthropogenic or Modern samples are actually determined by the $^{81}$Kr/$^{83}$Kr loading rates in the sample compared with those same loading rates in our daily reference gas measurements. This superratio (SR) is defined as
\begin{equation}
\label{SR}
	^{81}\text{Kr}_{\text{SR}} = \frac{^{81}\text{Kr}_{\text{Sample}}/^{83}\text{Kr}_{\text{Sample}}}{^{81}\text{Kr}_{\text{Reference}}/^{83}\text{Kr}_{\text{Reference}}}
\end{equation}
and gives us the $^{81}$Kr/Kr isotopic ratio with the systematic effects due to these drifts removed. Our standard reference gas is the same as our Modern sample, but it is crucial to note that there are differences in the way that we measure the reference from how we typically measure a sample. This is due to a cross-sample contamination problem, which we describe below (and which is discussed in \citet{J2012}).

The radio-frequency discharge which excites the krypton also ionizes a fraction of the atoms, and gives them high enough energy to implant themselves inside the surfaces of the source chamber. Because ATTA is used to measure very small amounts of krypton gas ($\sim\mu$L of krypton at STP), the gas must be recirculated in our vacuum system for several hours. If there are implanted atoms from previous samples present during the measurement of a new sample, they will be knocked out by the new ions and accumulate as a contaminant within the sample gas volume as the gas is continuously recycled. To avoid this problem during our reference measurements, we measure the standard reference gas without recirculation in order to prevent the accumulation of any significant contamination. We can do this because we have a plethora of reference gas in comparison to the amount of gas we typically have in a sample. However, since we have not previously studied the systematics below the 5\% level between the ``closed mode'' where the gas is recirculated, and the ``open mode'' where it is not, we cannot simply compare the PreAnthropogenic $^{81}$Kr/Kr isotopic ratio given by equation (\ref{SR}) to unity just because we use the Modern sample as our reference gas. Instead, we must follow the normal procedures for measuring both samples in closed mode in order to avoid any potential systematics from these two different modes of measurement. Fortunately, by doing so, we will also be probing for these very systematics we wish to avoid since the Modern gas will be measured as both sample and reference to 1\% precision.

The measurements of the samples are conducted in the following manner. 8 $\mu$L of krypton at STP of sample gas is injected into the vacuum system and recirculated in the closed mode configuration. The measurement then takes 2.5-4.5 hours. During the measurement, laser frequencies automatically switch between the different krypton isotopes and thus measure their various loading rates in five-minute cycles. 3.5 minutes is spent on $^{81}$Kr, 1 minute on $^{85}$Kr, and 0.5 minutes on $^{83}$Kr before the sequence repeats. We measure $^{85}$Kr to determine the cross-sample contamination since the $^{85}$Kr isotopic abundances of our samples are known through LLC.

Once the sample measurement is complete, the system is opened to turbo-molecular pumps which remove the sample from the system. Then, with the system left open to these pumps, i.e. the open mode configuration, krypton gas from the reference bottle is flowed through the system at a constant pressure. The loading rates for the krypton isotopes are then measured for this reference gas (which is the same gas as the Modern sample). This reference measurement requires about 2-3 hours. Once the measurement is complete, argon gas is then flowed through the system in the open mode configuration with the plasma discharge active overnight. The argon discharge removes the krypton atoms embedded in the source chamber. This argon ``washing'' procedure reduces the total amount of contaminant krypton that can outgas from the chamber walls during a measurement to $<$2\% of the sample gas being measured.

\section{Results}

\begin{figure}[t]
\includegraphics[width=30pc]{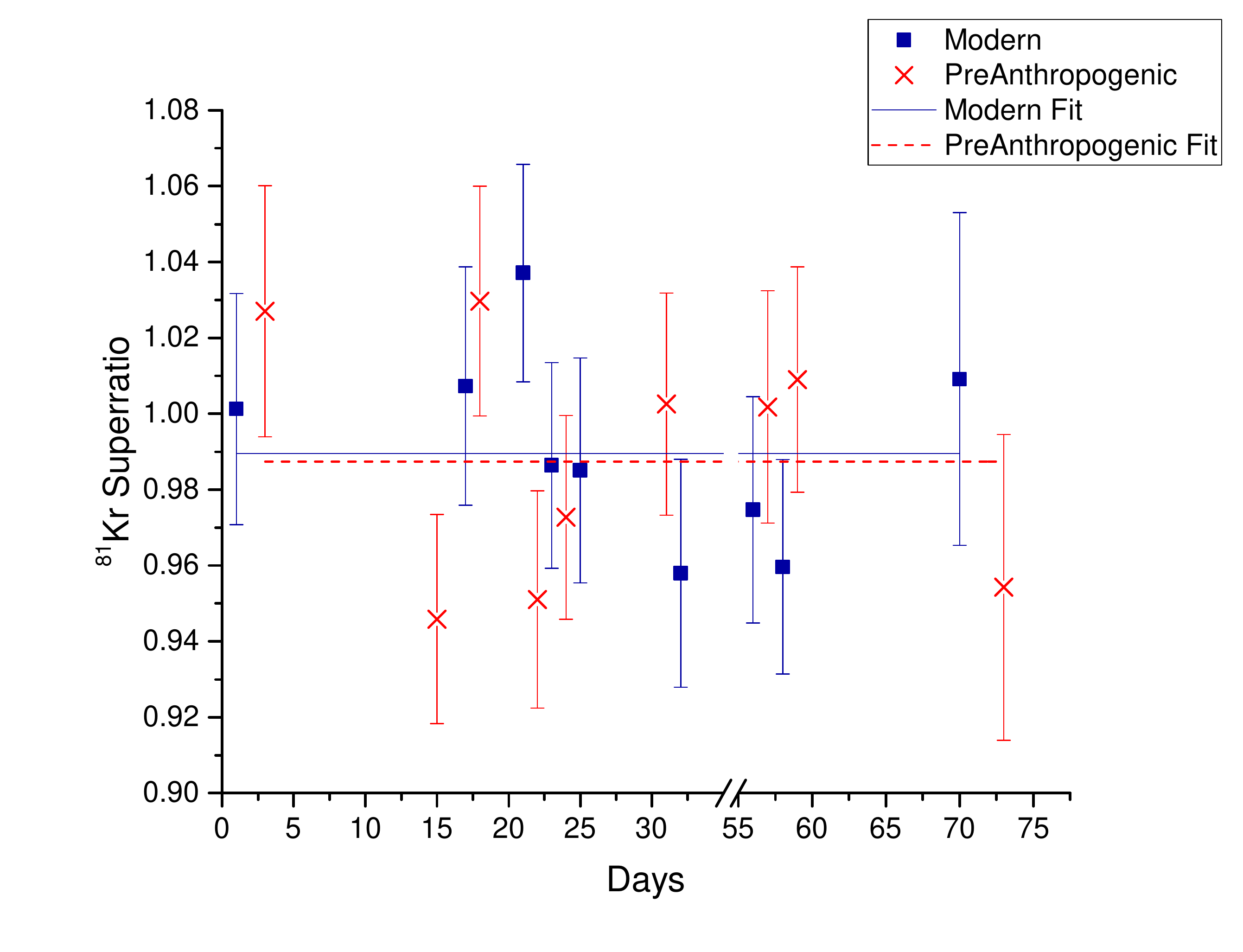}
\caption{\small{Superratios for $^{81}$Kr in both ``Modern'' (blue squares) and ``PreAnthropogenic'' (red crosses). \added{The error bars represent one-sigma uncertainties in each measurement.} The consistency of the results over the 75 day period demonstrates the robustness of our long-term systematic control.}}
\label{kr81}
\end{figure}

Measurements for this experiment were taken in the manner described in Section \ref{procedure} over the course of a two-and-a-half-month period. Nine measurements were conducted for each of the two samples to obtain sufficient statistical precision for a 1\% measurement. Major realignment of the laser system was done at several junctures throughout the period, but never in between sample and reference measurements. The results of these measurements are presented in Figure \ref{kr81}. Blue squares represent the Modern samples and red crosses represent the PreAnthropogenic samples. \added{The error bars represent one-sigma uncertainties in each measurement.} The superratios for both the Modern and PreAnthropogenic samples are plotted and fit to a constant function. The reduced chi-squared values of these fits are 0.73 and 1.16 for the Modern and PreAnthropogenic samples, respectively. The weighted average of the measurements for each sample type, using the single measurement errors as our weights, gives us
\begin{eqnarray}
	^{81}\text{Kr}_{\text{Modern, SR}} &=& 0.990 \pm 0.010 \nonumber \\
	^{81}\text{Kr}_{\text{PreAnthropogenic, SR}} &=& 0.988 \pm 0.010.\nonumber
\end{eqnarray}
\added{using one-sigma uncertainties.}

The two average values agree within error, meaning that we have no significant anthropogenic contribution. Consequently, we report a 2.5\% upper limit at the 90\% confidence level on anthropogenic contributions to the atmospheric abundance of $^{81}$Kr. The Modern value also represents good agreement between the open mode and closed mode measurements. As a check on the influence of cross-sample contamination mentioned earlier, we conduct outgassing tests on the stable isotopes of krypton prior to each run and determined a limit of 0.5\% contamination per hour, primarily due to gas from the reference bottle. Additionally, we have also compared the $^{85}$Kr loading rates from each measurement. On average the Modern $^{85}$Kr loading rate was 8740 atoms/hour and the PreAnthropogenic $^{85}$Kr loading rate was 150 atoms/hour. Presuming that all of the activity measured in the PreAnthropogenic sample was from the Modern gas suggests that we have a 1.8\% maximum contamination effect. Given the average runtime of 3.5 hours for the PreAnthropogenic runs, this result agrees with our outgassing test limit of just below 2\%.

This 2.5\% limit on anthropogenic $^{81}$Kr in the atmosphere is in good agreement with our theoretical model, which expects a $\leq$0.15\% anthropogenic contribution. In placing this limit we have demonstrated that the ATTA technique can measure samples down to 1\% precision provided sufficient statistics. We have also removed concerns of systematic effects from anthropogenic contributions when using $^{81}$Kr as an environmental tracer at this level of precision. 

\section{Acknowledgements}
\noindent We thank Zheng-Tian Lu for his guidance in the project and suggestions on this manuscript. This work is supported by Department of Energy, Office of Nuclear Physics, under Contract No. DEAC02-06CH11357. We also acknowledge funding from an Argonne/University of Chicago Collaborative Seed Grant.

\end{document}